\documentclass[conference]{IEEEtran}
\IEEEoverridecommandlockouts

\usepackage{cite}
\usepackage{amsmath,amssymb,amsfonts}
\usepackage{graphicx}
\usepackage{textcomp}
\usepackage{xcolor}
\def\BibTeX{{\rm B\kern-.05em{\sc i\kern-.025em b}\kern-.08em
    T\kern-.1667em\lower.7ex\hbox{E}\kern-.125emX}}

\usepackage[hyphens]{url}
\usepackage{algorithm}
\usepackage{algorithmicx}
\usepackage{algpseudocode}
\usepackage{verbatim} 
\usepackage{pifont}
\usepackage{multirow}
\usepackage{makecell}
\usepackage{enumerate}
\usepackage{flushend}
\usepackage{listings}
\usepackage{enumitem}
\usepackage[capitalise]{cleveref}

\crefformat{section}{\S#2#1#3}
\crefformat{equation}{Policy #2#1#3}

\lstdefinelanguage{x86-64}{
  alsoletter={.},
  morekeywords=[1]{ 
     cmpl, jne, je, lea, call, jmp, lea, movzbl, cmp, movb
   },
    morekeywords=[2]{ 
    .align, .ascii, .asciiz, .byte, .data, .double, .extern,
    .float, .globl, .half, .kdata, .ktext, .set, .space, .text, .word, SENSITIVE_BEGIN, SENSITIVE_END, .L1, .L2
  },
  morekeywords=[3]{ 
    rbp, rip, rdi, al
  },
  keywordstyle=[2]\color{black!80!black},      
  keywordstyle=[3]\color{black!50!black},         
  sensitive=true,  
  morecomment=[l]{;},  
  morestring=[b]",   
  morestring=[b]'    
}

\newcommand{\keypoint}[1]{\noindent\textbf{#1}}

\begin{document}
%
\title{PhantomFetch: Obfuscating Loads against Prefetcher Side-Channel Attacks}

\author{
    \IEEEauthorblockN{Xingzhi Zhang\IEEEauthorrefmark{1}}
    \IEEEauthorblockA{
        Zhejiang University \\
        xingzhizhang@zju.edu.cn
        \thanks{\IEEEauthorrefmark{1}Co-first authors: these authors contributed equally to this work.}
    }
    \and
    \IEEEauthorblockN{Buyi Lv\IEEEauthorrefmark{1}}
    \IEEEauthorblockA{
        Zhejiang University \\
        buyilv@zju.edu.cn
    }
    \and
    \IEEEauthorblockN{Yimin Lu}
    \IEEEauthorblockA{
        Zhejiang University \\
        lu\_yimin@zju.edu.cn
    }
    \and
    \IEEEauthorblockN{Kai Bu}
    \IEEEauthorblockA{
        Zhejiang University \\
        kaibu@zju.edu.cn
    }
}

\maketitle



%


\begin{abstract}
The IP-stride prefetcher has recently been exploited to leak secrets through side-channel attacks. It, however, cannot be simply disabled for security with prefetching speedup as a sacrifice. The state-of-the-art defense tries to retain the prefetching effect by hardware modification. In this paper, we present PhantomFetch as the first prefetching-retentive and hardware-agnostic defense. It avoids potential remanufacturing cost and enriches applicability to off-the-shelf devices. The key idea is to directly break the exploitable coupling between trained prefetcher entries and the victim's secret-dependent loads by obfuscating the sensitive load effects of the victim. The experiment results show that PhantomFetch can secure the IP-stride prefetcher with only negligible overhead.
\end{abstract}


\section{Introduction}

Prefetcher side-channel attacks \cite{gruss2016prefetch, shin2018unveiling, cronin2019fetching, rohan2020reverse, vicarte2022augury, lipp2022amd} have recently evolved to leak secrets without relying on primitives for cache side channels \cite{chen2023afterimage}. Such attacks exploit the IP-stride prefetcher on modern processors. This prefetcher provides performance speedup by \ding{182} learning repetitive strides between memory addresses requested by load instructions with identical least significant bits in instruction pointers (IPs), \ding{183} prefetching data addressed by the sum of the learned stride to the current requested address, and \ding{184} avoiding potential cache misses of the prefetched data \cite{IntelManual}. All processes on the same core can share the IP-stride prefetcher, opening the door for side-channel attacks. Specifically, the attacker process trains prefetcher entries using loads that have the identical least significant bits in IPs with that of the secret-dependent loads by the victim process \cite{chen2023afterimage}. Then any secret-dependent load from the victim may prefetch data in the attacker process' address space. Ultimately, the attacker process can correctly infer the victim's secret based on the fact that the prefetched data can already be hit in the cache upon the first request by the attacker process.

The existing defenses, however, fail to simultaneously satisfy prefetching retention and hardware agnosticism. For example, a line of defenses simply disables the prefetcher \cite{boran2021disabling, chen2023afterimage, schluter2024scheduling}. This is absolutely effective against any side-channel attack that exploits prefetching. However, the IP-stride prefetcher may be simply disabled in practice because prefetching can yield up to 98\% speedup \cite{chen1994performance, vanderwiel2000data, srinath2007feedback}. The latest defense \cite{chen2023afterimage} modifies hardware logic to flush prefetcher entries upon context switches such that trained prefetching effect is obscured across processes. However, the required hardware modification may lead to diseconomy of remanufacturing cost and inapplicability to off-the-shelf devices.

In this paper, we present PhantomFetch suites as the first prefetching-retentive and hardware-agnostic defense against IP-stride--prefetcher side-channel attacks. The key idea is to obfuscate the impact of the victim's secret-dependent loads on prefetching effect. We explore two run-time load obfuscation schemes. One is load injection, which injects a set of well-crafted loads to erase trained prefetcher entries by the victim before any other process can access and exploit these entries. The other is load relocation, which randomly swaps locations of the victim's secret-dependent loads such that the same trained prefetcher entry could couple with any secret-dependent load.

We design and implement two variants, PhantomFetch-vLI and PhantomFetch-vLR, to practice load injection and load relocation, respectively. PhantomFetch-vLI modifies kernel function \texttt{context\_switch} by leading it with non-preemptive execution of our crafted loads. We leverage the fact that load IPs with the same 8 least significant bits match the same prefetcher entry \cite{chen2023afterimage} to craft loads for injection. A two-round load injection strategy is developed to ensure the complete obscurity of prefetcher entries trained by the victim upon context switches. Furthermore, PhantomFetch-vLR provides a compiler-oriented alternative for scenarios where kernel modification might not be permitted. It enables the compiler to instrument the victim program with a load-relocation gadget. The instrumented program can extract and relocate secret-dependent if-blocks and else-blocks, through a randomization process, at run time.

In summary, we make the following major contributions to secure IP-stride prefetcher without hardware modification.
\begin{itemize}
    \item We investigate the state-of-the-art defenses against IP-stride--prefetcher side-channel attacks (Section~\ref{sec:problem}). We identify their fundamental lack of simultaneous satisfaction of prefetching retention and hardware agnosticism. This prevents them from retaining the prefetching effect while readily applying it to off-the-shelf devices.
    \item We present PhantomFetch-vLI and PhantomFetch-vLR as the first prefetching-retentive and hardware-agnostic defenses (Section~\ref{sec:overview}). Their core idea is load obfuscation, which can obscure exploitable secret-dependent prefetching effect trained by the victim at run time.
    \item We implement and evaluate PhantomFetch-vLI and PhantomFetch-vLR with representative test cases (Section~\ref{sec:evaluation}). The results show that PhantomFetch-vLI and PhantomFetch-vLR can secure the IP-stride prefetcher with only 0.6\% slowdown and 4.0\% slowdown, respectively.
\end{itemize}



\section{Problem}
\label{sec:problem}

In this section, we first review the latest prefetcher side-channel attack that no longer relies on primitives for traditional cache side channels. We then investigate existing defenses and identify their limitation on simultaneously attaining both prefetching retention and hardware agnosticism.

\subsection{Prefetcher Side-Channel Attack}
\label{subsec:attack}

Most existing prefetcher side-channel attacks \cite{gruss2016prefetch, shin2018unveiling, cronin2019fetching, rohan2020reverse, vicarte2022augury, lipp2022amd} still have to rely on cache side channels to ultimately induce information leakage \cite{chen2023afterimage}. A common rationale of them exploits the cache-status variation by prefetched data. Such cache-status variation leaks information through various cache side-channel attacks \cite{yarom2014flush+, gruss2016flush+, gruss2015cache, liu2015last, osvik2006cache}.
However, after almost 20 years of the emergence of cache side-channel attacks \cite{osvik2006cache}, a plethora of secure cache designs have already been practiced \cite{li2024treasurecache}. The closing of cache side channels further throttles prefetcher side channels that build upon them otherwise.

\begin{figure}[t]
    \centering
    \includegraphics[width=\linewidth]{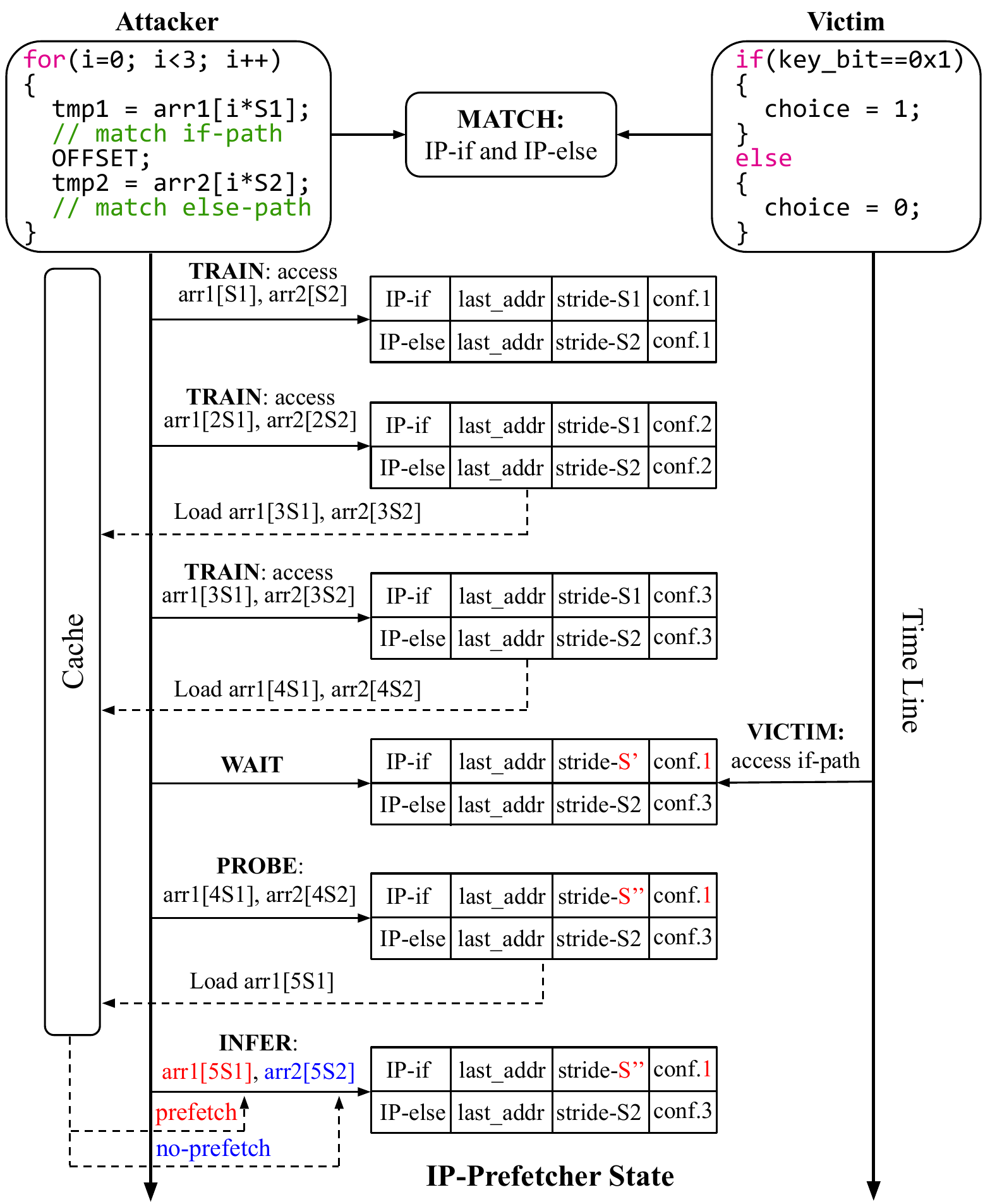}
    \vspace{-6mm}
    \caption{AfterImage attack flow by exploiting the IP-stride prefetcher to infer the secrets of the victim.
    }
    \label{fig:attack}
    \vspace{-5mm}
\end{figure}

The latest prefetcher side-channel attack can leak information even if primitives for cache side channels are prohibited. One such representative is the AfterImage attack \cite{chen2023afterimage} (Figure~\ref{fig:attack}). It infers the victim's secret from how the victim's secret-dependent load operations affect the IP-stride prefetcher, which further affects the attacker's own cache access trace. It does not depend on the specific cache access trace of the victim as in cache side-channel attacks. 

\noindent \textbf{Step 1: Train prefetcher.} The attacker starts with training certain prefetcher entries such that their IP fields can match the addresses of secret-dependent loads by the victim. Specifically, the IP field of a prefetcher entry matches only the least significant 8 bits of a load instruction's address. This greatly eases identification of the IPs of exploitable loads. Even if the victim does not belong to a shared library\footnote{When the victim code belongs to a shared library, the attacker can determine the IPs of loads therein using disassembly tools such as \texttt{objdump}.}, the attacker only needs to exercise up to 256 possibilities for quickly determining the exact lowest 8 bits of a target load \cite{chen2023afterimage}. The attacker then trains two prefetcher entries in particular---one matches a load from the if-path of the victim code and the other matches another load from the else-path of the victim code. The stride field in a prefetcher entry records the distance between the memory addresses of the data requested in two consecutive loads. The prefetcher will update the value of the stride field to the value of $\mathit{current\_address-last\_address}$. For ease of computing such distance, each entry contains also a $\mathit{last\_address}$ field to record the requested address by the most recent load. If a load satisfies both IP matching and stride equivalence, the confidence field in the matched entry is incremented (with 3 as a saturated state). Once the confidence reaches 2, the current load with the IP field matched triggers prefetching. The address of data to prefetch is the sum of the stride value and the requested address of the current load.

\noindent \textbf{Step 2: Wait for secret-dependent load.} After training exploitable prefetcher entries, the attacker waits for secret-dependent loads to take place by the victim. We use the case when the victim executes the if-path for example in Figure~\ref{fig:attack}. The executed load matches the IP-if entry and also triggers prefetching because the confidence is no less than 2. Of particular interest to the attacker is how this load impacts effectiveness of the matched IP-if entry. Specifically, the load by the victim may not exactly follow the same stride of $S1$ trained by the attacker. The stride field is accordingly updated to $S'$, which is equal to the distance between the victim's requested address and the $\mathit{last\_address}$ in the entry. The confidence is then reset as 1. This turns the entry ineffective for prefetching.

\noindent \textbf{Step 3: Probe prefetcher.} The attacker then probes the trained prefetcher entries using the same set of loads in Step 1. The corresponding requested addresses are $4S1$ and $4S2$ in this example. If both trained prefetcher entries are still effective, they should be triggered to prefetch data with addresses $5S1$ and $5S2$ in advance.

\noindent \textbf{Step 4: Infer secret.} Finally, the attacker probes the supposed prefetched data to infer the victim's secret. Since the IP-if entry becomes ineffective due to the victim's if-path execution, data addressed $5S1$ cannot be prefetched. With the cache miss of $5S1$-addressed data can the attacker infer the if-path execution and the associated secret of the victim.

\subsection{Defenses and Limitations}
We find that none of the existing defenses against AfterImage-alike prefetcher side-channel attacks can simultaneously satisfy prefetching retention and hardware agnosticism. Prefetching retention requires that a defense retain the functionality of the prefetcher. This ensures that speedup by prefetching is preserved at best. Hardware agnosticism requires that a defense need not necessarily modify hardware logic. This minimizes remanufacturing cost and maximizes solution portability, especially for off-the-shelf devices. 

\noindent \textbf{Diasble prefetcher.} The first defense to resort to would be simply disabling the prefetcher. Once the prefetcher is disabled, any side channel built upon it becomes impossible \cite{boran2021disabling, chen2023afterimage, schluter2024scheduling}. With the closing of prefetcher side channels comes the inevitable sacrifice of performance speedup that the prefetcher can offer. A recent work disables the prefetcher for only security-critical code \cite{schluter2024scheduling}. However, security-critical code is not necessarily less performance-critical. Disabling the prefetcher for only security-critical code may still lead to an unacceptable performance slowdown. A more efficient defense should retain the prefetching effect as much as possible.

\noindent \textbf{Flush prefetcher entries.} The hardware solution that flushes prefetcher entries upon context switches \cite{chen2023afterimage} aims exactly at a prefetching-retentive defense. Its key motivation is that not every prefetcher access is vulnerable to AfterImage-alike side-channel attacks. Only prefetcher accesses right after context switches can be exploited (Section~\ref{subsec:attack}). Therefore, the solution in \cite{chen2023afterimage} flushes all entries in the IP-stride prefetcher upon context switches. This leaves the attacker with no established prefetcher entries to be exploited afterward. Between two context switches, the prefetcher can still be trained by a process from scratch and then unleash the prefetching effect again through well-trained entries. Experiment results show that this solution introduces only a negligible performance slowdown. Albeit guaranteeing efficiency with prefetching retention, the flush-prefetcher solution needs to modify hardware. Such a lack of hardware agnosticism leads to potential manufacturing cost and limits its portability to off-the-shelf devices.

\section{PhantomFetch}
\label{sec:overview}

In this section, we present PhantomFetch suites as the first prefetch-retentive and hardware-agnostic defense against prefetcher side-channel attacks (Table~\ref{tab:property}). The key idea is load obfuscation through which exploitable load effects are obfuscated and thus prefetcher side-channel attacks are prevented. We explore two types of obfuscation schemes in particular---OS-based load injection and compiler-based load relocation. They are tailored for different adoption scenarios with either OS or compiler being more convenient to modify. Load injection enables the OS to inject crafted load instructions to invalidate trained prefetcher entries upon context switching. Load relocation enables the compiler to instrument secret-dependent branches such that if-block and else-block can run at each other's locations. Both schemes can successfully decouple the exploitable mapping between load instructions and their addresses. 

\begin{table}[t]
\caption{Property comparison of defenses against prefetcher side-channel attacks.}
\begin{center}
\begin{tabular}{|c|c|c|}
\hline
\multirow{3}{*}{\textbf{Defense}}&\multicolumn{2}{c|}{\textbf{Property}} \\
\cline{2-3} 
 & \textbf{Prefetching}& \textbf{Hardware} \\
& \textbf{Retention} & \textbf{Agnosticism} \\
\hline
Disable Prefetcher & \multirow{2}{*}{\ding{55}} & \multirow{2}{*}{\ding{51}} \\
for All or Selective Code & & \\
\hline
Flush Prefetcher Entries & \multirow{2}{*}{\ding{51}} & \multirow{2}{*}{\ding{55}} \\
upon Context Switch & & \\
\hline
Obfuscate Load & \multirow{2}{*}{\ding{51}} & \multirow{2}{*}{\ding{51}} \\
(PhantomFetch in this paper) & & \\
\hline
\end{tabular}
\label{tab:property}
\end{center}
\vspace{-6mm}
\end{table}






\subsection{PhantomFetch via Load Injection: PhantomFetch-vLI}

The key OS modification by PhantomFetch-vLI lies in the kernel function---\texttt{context\_switch}. Since PhantomFetch-vLI strives for a software alternative solution for the hardware flush-prefetcher defense, the injected loads during a context switch should replace all the prefetcher entries before the prefetcher is accessed by the next process. This effectively erases any deliberately trained prefetcher status by a potential attacking process prior to the context switch. Potential prefetcher side-channel attacks are thus prevented as well. To achieve this goal, we need to address the following critical design questions for implementing PhantomFetch-vLI. 
\begin{itemize}
    \item When to inject load instructions during context switches? This decides whether the injection can take effect and to what extent it impacts context-switch speed.
    \item What specific load instructions to inject? This decides whether an injected load instruction can erase a target trained prefetcher entry.
    \item How to inject the crafted load instructions? This determines if the load scheduling can effectively ensure the complete obscurity of prefetcher entries.
\end{itemize}

\noindent \keypoint{When to inject: Non-preemptive load injection at the beginning of \texttt{context\_switch}.} Two design choices are adopted to address the first design question. First, the injected loads are executed in a non-preemptive way. This prevents load injection from being interrupted by, for example, context switching to the attacker's process. Second, we run load injection immediately at the beginning of \texttt{context\_switch}. Since AfterImage takes effect after a context switch, we do not need to constantly enforce load injection to obscure prefetcher entries. Injecting loads only when \texttt{context\_switch} starts not only zeros the attack window but also minimizes the corresponding performance overhead.

\noindent \textbf{What to inject: Load instructions with addresses having different lowest 8 bits.} This design choice leverages the fact that the IP-stride prefetcher is indexed by the least significant 8-bits of the addresses of load instructions \cite{chen2023afterimage}. Any two load instructions with the same lowest 8 bits will be mapped to the same prefetcher entry. Therefore, to minimize the number of load instructions to inject, we require that their addresses have different lowest 8 bits. 

\noindent \textbf{How to inject: Two-round load injection toward a complete erasure of trained prefetcher entries.} How to inject crafted loads depends on the IPs of trained prefetcher entries and the prefetcher's replacement policy. The IPs of trained prefetcher entries by the attacker are highly likely unknown. We thus cannot precisely determine the addresses of crafted load instructions. We address this challenge by leveraging the fact that the IP-stride prefetcher uses an LRU-alike replacement policy \cite{chen2023afterimage}. Injecting a sufficient number of crafted load instructions helps to replace all the prefetcher entries. The efficiency of this design choice is guaranteed by another prefetcher feature, that is, the IP-stride prefetcher contains only 24 entries \cite{chen2023afterimage}. The ideal case for PhantomFetch-vLI to fully obscure trained prefetcher entries injects only 24 consecutive load instructions with different lowest 8 bits among their addresses.

However, a single round of load injection cannot guarantee that all trained prefetcher entries are erased. Given the unknown statuses of trained prefetcher entries, an injected load instruction may happen to match and trigger a trained prefetcher entry. Let the address of this specific injected load instruction be denoted as $\mathit{load\_address}$. Then the aforementioned entry-triggering coincidence happens when two conditions are satisfied (Figure~\ref{fig:attack}). First, $\mathit{load\_address}$ and the IP of the triggered entry share the same lowest 8 bits. Second, $\mathit{load\_address}$ and two fields in the triggered entry---$\mathit{last\_address}$ and $\mathit{stride}$---satisfy the equation of $\mathit{load\_address} - \mathit{last\_address} = \mathit{stride}$. The triggered entry still retains its effectiveness trained by the potential attacker.

\begin{figure}[t]
    \centering
    \includegraphics[width=\linewidth]{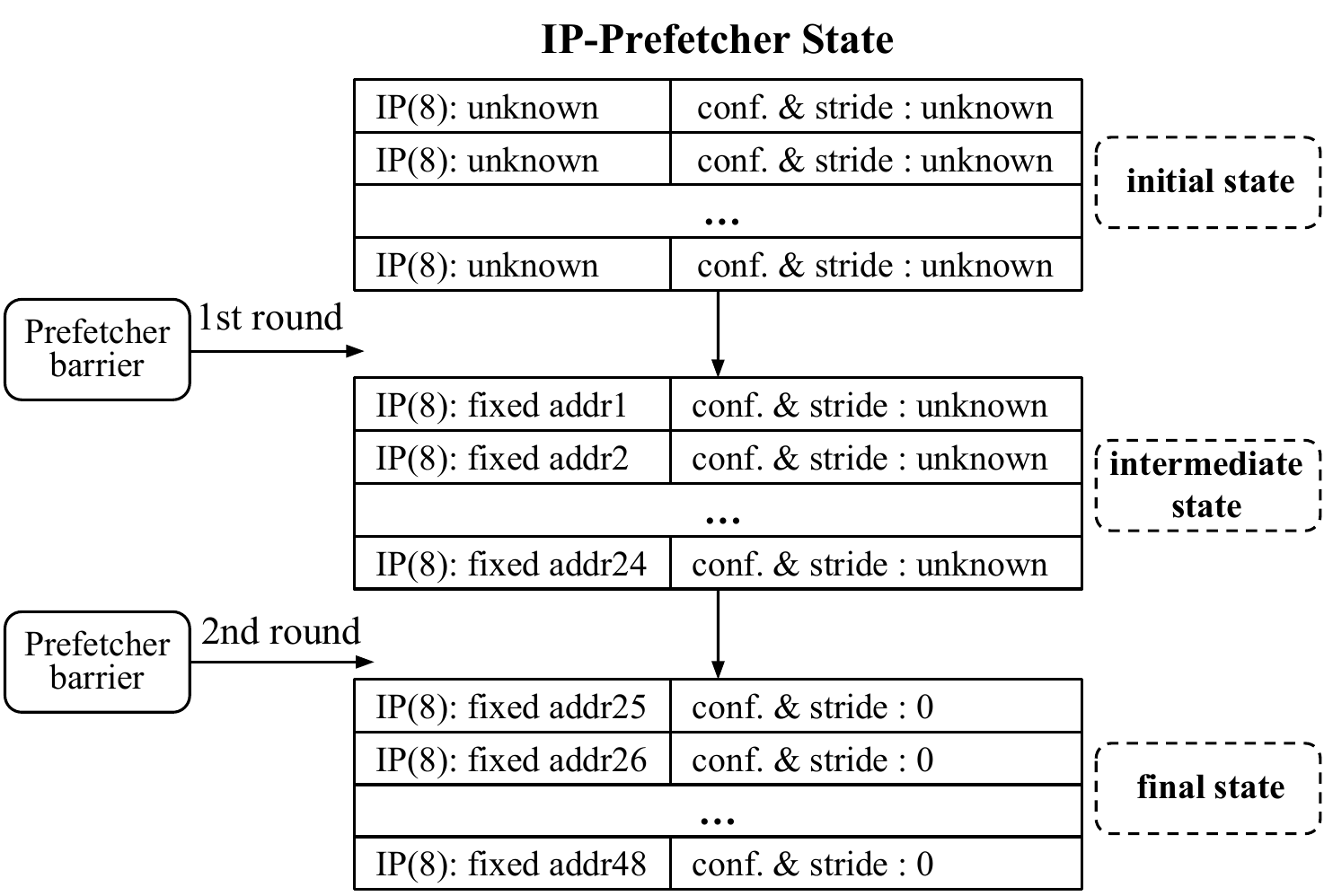}
    \vspace{-6mm}
    \caption{Two-round load injection toward complete obscurity of trained prefetcher entries.}
    \vspace{-5mm}
    \label{fig:injection}
\end{figure}

We explore a two-round load injection scheme to completely erase trained prefetcher entries. Figure~\ref{fig:injection} shows the framework. Targeting the 24-entry IP-stride prefetcher, we inject 24 loads and 24 loads for the first round and the second round, respectively. All the 48 load instructions have different lowest 8 bits in their addresses. After the first round, even if the 24 injected loads happen to trigger certain trained prefetcher entries, their addresses decide exactly the IPs of all the 24 prefetcher entries. This essentially turns the unknown IPs of trained prefetcher entries into known ones. In the second round, we continue to inject 24 other load instructions with different addresses than the known IPs. Each of these loads ensures to replace one of the known IPs' corresponding prefetcher entries. Once the second round completes, all the original trained prefetcher entries are replaced. Any potential information-leakage channel over them is eventually precluded.

\noindent \textbf{PhantomFetch-vLI implementation.} Finally, we implement PhantomFetch-vLI by introducing $\texttt{load\_injection}$ in a non-preemptive way at the beginning of \texttt{context\_switch}. 
The implementation of $\texttt{load\_injection}$ is detailed in Listing~1. It consecutively executes 48 load instructions with addresses that differ in their lowest 8 bits. Note that the prefetcher may take effect only if memory addresses requested by the load instructions have their page addresses translated in the TLB \cite{chen2023afterimage}. To cater to this potential implicit requirement, we confine the memory addresses requested by the injected loads within a single page. Only after $\texttt{load\_injection}$ completes to erase all the trained prefetcher entries prior to context switching can the functions in the original \texttt{context\_switch} be executed.

\begin{lstlisting}[caption={Load injection introuced in \texttt{context\_switch}.}, label={vli}, numbers=left, numberstyle=\tiny]
__attribute__((aligned(256)))
void load_injection(){
    asm volatile(
        "movq  %0,      %%rsi\n"
        "movq  (%%rsi), %%rax\n"
        ".REPT 47\n"
        "addq  $64,     %%rsi\n"
        "movq  (%%rsi), %%rax\n"
        ".ENDR \n"
        :: "r"(array)
        :  "%rax","rsi"
    );
}
\end{lstlisting}
\vspace{-3mm}


%


\subsection{PhantomFetch via Load Relocation: PhantomFetch-vLR}
%
Considering that it may take various considerations into account for an OS to update a kernel function like $\texttt{context\_switch}$, we further explore a compiler-based solution called PhantomFetch-vLR. It ensures that an executable program has already integrated the protection scheme. Prior to the launching of a secure OS, PhantomFetch-vLR can be adopted by communities of security-critical programming languages (e.g., C for commonly used cryptographic libraries such as Libgcrypt and OpenSSL). PhantomFetch-vLR can also function as a stand-alone tool for instrumenting a program. Then the instrumented program is fed to the unmodified compiler. However, this design choice prolongs the software development process. Software developers may also trust a third-party tool less than the prevailing compiler. We thus choose to incorporate PhantomFetch-vLR into the compiler.

\begin{figure}[t]
    \centering
    \includegraphics[width=\linewidth]{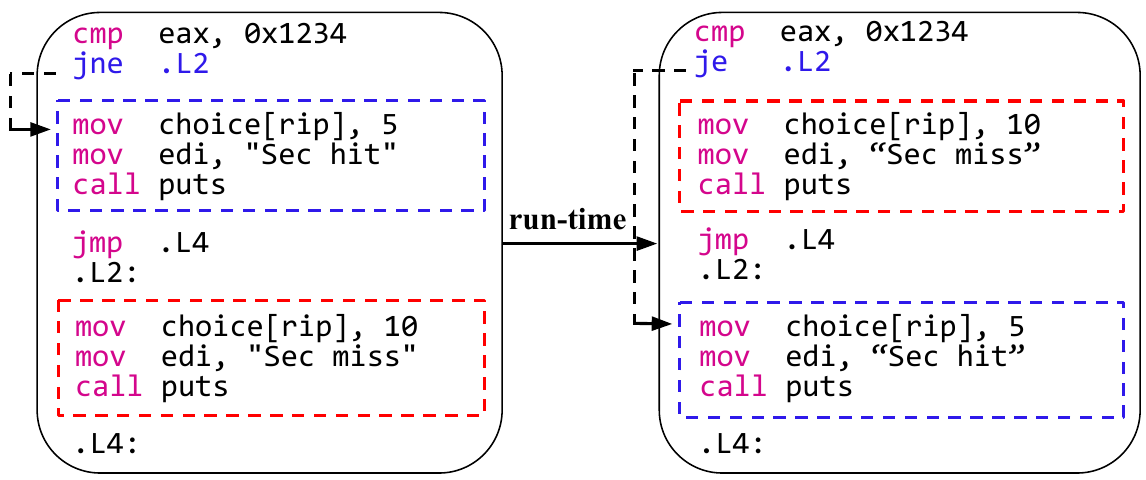}
    \vspace{-6mm}
    \caption{Example of load-relocation effect.}
    \vspace{-6mm}
    \label{fig:relocation}
\end{figure}

The key idea of PhantomFetch-vLR is obfuscating control flow through load relocation. In the AfterImage-alike attacks, the attacker needs to correctly identify which path's load instructions of a target \texttt{if-else} branch are executed by the victim. The identification leverages the known mapping policy of load-instruction addresses and prefetcher entries. Load relocation aims to obfuscate the mapping process by dynamically adjusting the locations and thus addresses of such load instructions. The condition-test instruction may be re-encoded meanwhile to preserve program correctness. Figure~\ref{fig:relocation} shows the run-time obfuscation effect. The example code essentially takes the if-path (located at \texttt{mov choice[rip], 5}) if the value in register \texttt{eax} is equal to 0x1234 and takes the else-path (located at \texttt{mov choice[rip], 10}) otherwise. With run-time load relocation, locations of the two paths might be swapped as illustrated. This leaves the attacker with different addresses to be monitored even if the same set of load instructions are executed.

\begin{lstlisting}[caption={Gadget for invoking load relocation.},label={gadget}]
// GADGET inserted by compiler
GADGET:
if(read_cyc() % 2)
   obfuscation(SENSITIVE_BEGIN, SENSITIVE_END, 1);
else
   obfuscation(SENSITIVE_BEGIN, SENSITIVE_END, 0);
SENSITIVE_BEGIN:
// secret-dependent branch
SENSITIVE_END:
\end{lstlisting}

\begin{figure}[t]
    \centering
    \includegraphics[width=\linewidth]{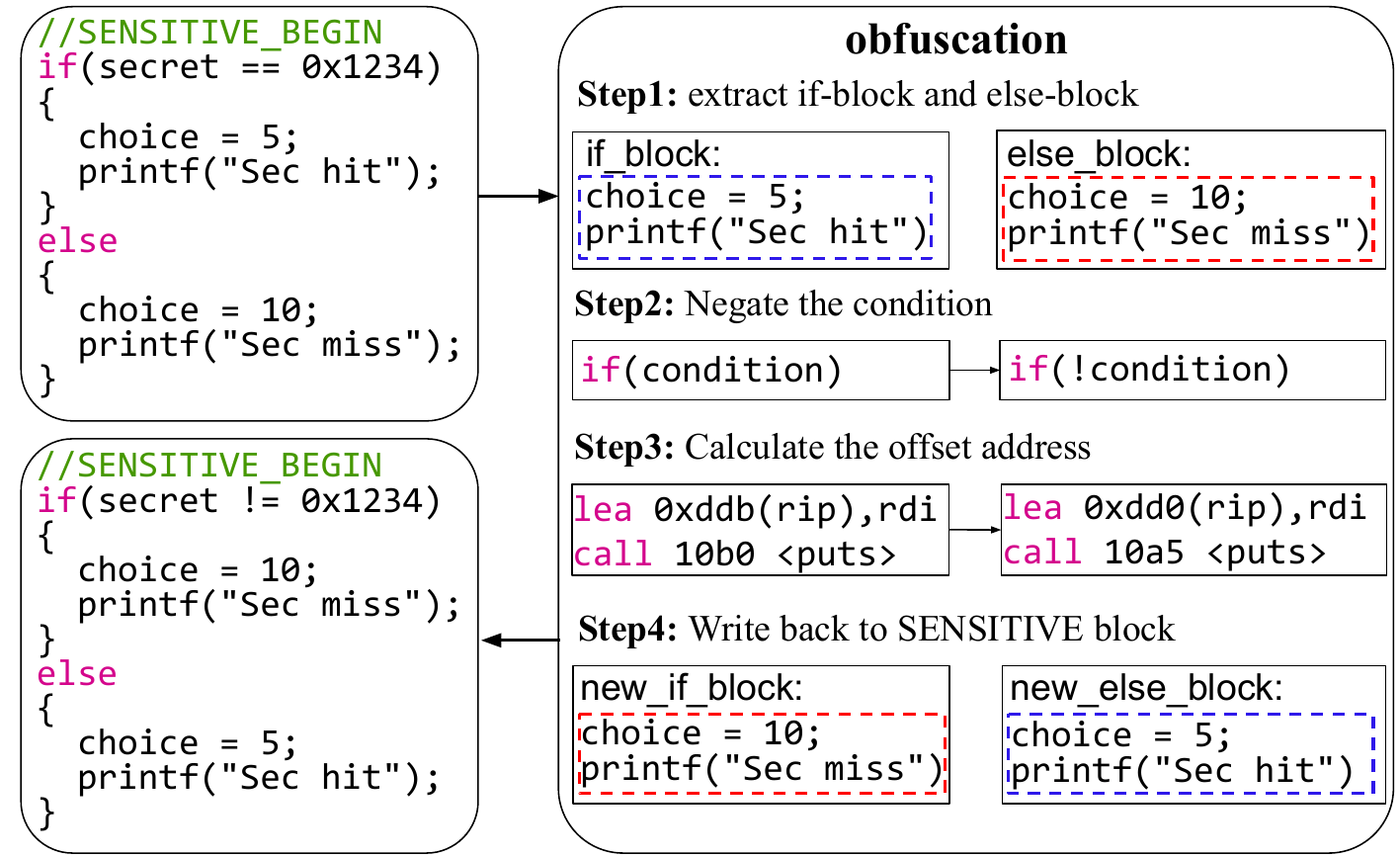}
    \vspace{-6mm}
    \caption{PhantomFetch-vLR framework.}
    \vspace{-6mm}
    \label{fig:PhantomC steps}
\end{figure}

\noindent \textbf{PhantomFetch-vLR implementation.} Figure~\ref{fig:PhantomC steps} highlights the framework for PhantomFetch-vLR to enforce run-time load relocation. For ease of presentation, we denote the start and the end of an \texttt{if-else} block as \texttt{SENSITIVE\_BEGIN} and \texttt{SENSITIVE\_END}, respectively. PhantomFetch-vLR enables the compiler to insert a gadget before \texttt{SENSITIVE\_BEGIN} (\cref{gadget}). The gadget randomly calls either instantiation of the \texttt{obfuscation} function with different parameters, which determine whether the subsequent \texttt{if-else} block should be actually obfuscated with load relocation or not. In \cref{gadget}, we directly use the value in the cycle register for randomizing the selection of \texttt{obfuscation} instances. The instance of \texttt{obfuscation(SENSITIVE\_BEGIN, SENSITIVE\_END, 1)} enforces load relocation. We use registers to pass parameters while calling the \texttt{obfuscation} function; no memory access is involved. This prevents the attacker from monitoring which path of \texttt{GADGET} is executed. Furthermore, \texttt{obfuscation} is executed regardless of which path of \texttt{GADGET} is executed, the attacker cannot tell if the \texttt{if-else} block is actually obfuscated by simply observing that whether \texttt{obfuscation} is executed or not.

\begin{lstlisting}[language=x86-64, caption={Dumped assembly code of secret-dependent branch.},label={dump}]
0x1151: <SENSITIVE_BEGIN>:
0x1151:	movzbl 0x2eba(%rip), %eax
0x1158:	cmp    $0xa, %al
0x115a:	jne    1171 <.L1>
0x115c:	movb   $0x5, 0x2eae(%rip) #4011 <choice>
0x1163:	lea    0xe9a(%rip),%rdi   #2004 "Sec hit"
0x116a:	call   1050 <puts@plt>
0x116f:	jmp    1184 <SENSITIVE_END>
0x1171: <.L1>:
0x1171:	movb   $0xa, 0x2e99(%rip) #4011 <choice>
0x1178:	lea    0xe90(%rip),%rdi   #200f "Sec miss"
0x117f:	call   1050 <puts@plt>
0x1184: <SENSITIVE_END>:
\end{lstlisting}

We showcase the design of \texttt{obfuscation} using the case in Figure~\ref{fig:PhantomC steps} when it alters the contents of the instruction area and swaps the if-path and else-path yet still preserves program semantics. It features the following four steps for such run-time load relocation. All these steps take effect over the \texttt{if-else} block's assembly code dumped as \cref{dump}.

\noindent \textbf{Step 1: Extraction of if-block and else-block.} The \texttt{obfuscation} function first identifies and extracts if-block and else-block of a branch labeled \texttt{SENSITIVE\_BEGIN} and \texttt{SENSITIVE\_END}. Specifically, the if-block and else-block are delineated by the instruction that jumps to \texttt{SENSITIVE\_END} (line 8). This instruction marks the end of the if-block and the beginning of the else-block. Furthermore, the if-block starts with the instruction next to the adjacent \texttt{cmp} and \texttt{jne}; the else-block ends with \texttt{SENSITIVE\_END}.

\noindent \textbf{Step 2: Negation of branch condition.} Since the jump instruction (line 4) determines control flow, its condition should be negated to preserve program semantics after swapping if-block and else-block. Its jump distance should also be adjusted to the length of the else-block. Modifying the jump instruction in this step and relocating instructions afterward requires writing permission over the page(s) containing the \texttt{if-else} block. We acquire writing permission using \texttt{mprotect}.

\noindent \textbf{Step 3: Calculation of offset address.} We continue to calculate the addresses of instructions for to-be-swapped if-block and else-block. This step applies to instructions using IP-relative addressing in particular. IP-relative addressing calculates the next fetch address based on the current value of IP register \texttt{rip} plus an offset value.  
As shown in line 5 of \cref{dump}, \texttt{movb} in the if-block adds offset \texttt{0x2eae} (encoded in the example binary machine code) to the current \texttt{rip} value as the target load address. When we relocate \texttt{movb} to the new-else block, the \texttt{rip} value would be different. We thus need to update the offset in \texttt{movb} encoding for the correct execution of \texttt{movb} in the new-else-block. In this example, we recalculate the offset of relocated \texttt{movb} to \texttt{0x4011} minus \texttt{0x1178}.

\noindent \textbf{Step 4: Relocation of if-block and else-block.} Finally, we complete \texttt{obfuscation} by physically swapping the if-block and else-block in memory. The original if-block with recalculated offset address is moved to the new-else-block (followed by \texttt{jmp SENSITIVE\_END}), which will be executed if the re-encoded branch in Step 2 fails the condition test. Similarly, the original else-block is moved to the new-if-block, which will be executed if the re-encoded branch is taken.

\section{Performance}
\label{sec:evaluation}



In this section, we evaluate the performance of PhantomFetch suites on a server with a 3.4 GHz CPU and 128 GB memory. Experiment results show that PhantomFetch-vLI and PhantomFetch-vLR introduce negligible performance overhead by 0.6\% slowdown and 4.0\% slowdown, respectively.









\subsection{PhantomFetch-vLI}

\begin{figure}[t]
    \centering
    \includegraphics[width=\linewidth]{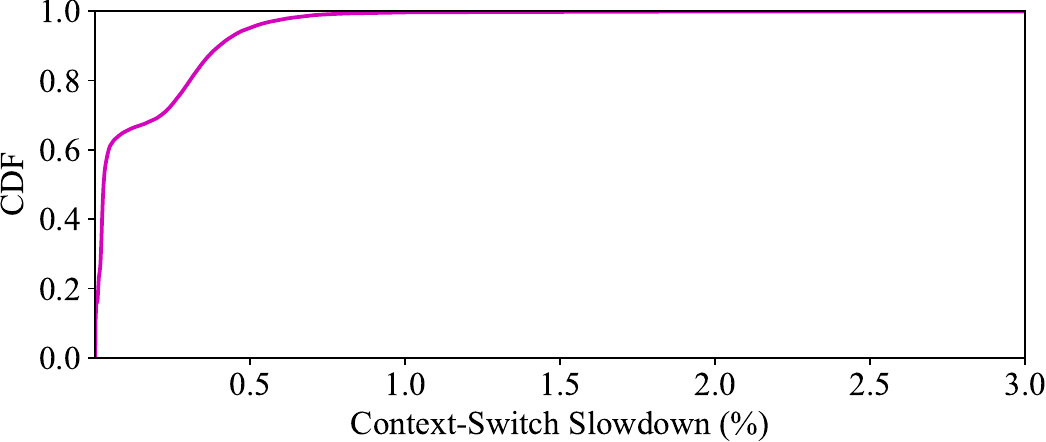}
    \vspace{-6mm}
    \caption{PhantomFetch-vLI overhead to context switching.}
    \vspace{-6mm}
    \label{fig:vli_cdf}
\end{figure}

We start with evaluating the overhead introduced by the injected loads per context switch. We implement PhantomFetch-vLI using Linux kernel v6.11.7 by modifying the \texttt{contex\_switch()} function in the file named \texttt{kernel/sched/core.c} \cite{Linux-kernel-code}. The context-switch overhead is defined by the ratio of the time for executing the injected loads to the overall execution time of the modified \texttt{contex\_switch()}. The \texttt{rdtsc} instruction is used for accurate timing estimation \cite{fang2015measuring}. We collect the timings over 100,000 \texttt{contex\_switch()} instances from kernel booting and program execution after the system is booted. In the Linux kernel boot process, context switches are primarily triggered by events such as page faults and I/O waits. Moreover, during program execution, context switches may be triggered by events such as system calls (e.g., \texttt{fork()}, \texttt{exec()}) and time-slice expirations. Both scenarios cover the majority of context-switch cases and thus suffice to generate representative timing samples. Figure~\ref{fig:vli_cdf} reports the context-switch overhead by PhantomFetch-vLI. 97.6\% of the overhead measurements can be as negligible as less than 0.6\%.

We further investigate the overhead in terms of slowdown. A generic estimation function will be formulated; we thus do not base our evaluation on sample benchmarks. Let $switch_{time}$ and $nonswitch_{time}$ denote the time cost by context switches and the remaining part of program execution time, respectively. The original program execution time prior to PhantomFetch-vLI adoption can be estimated as the following.
\begin{align}
program_{time}=switch_{time} + nonswitch_{time}.\nonumber
\end{align}
After using PhantomFetch-vLI, it introduces overhead to context switches (Figure~\ref{fig:vli_cdf}) and non-switch operations due to erasing trained prefetcher entries. Let such overhead be respectively denoted as $switch_{overhead}$ and $nonswitch_{overhead}$. The program execution time after PhantomFetch-vLI adoption is estimated as the following.
\vspace{-2mm}
\begin{align}
    program_{time'}=switch_{time}\times(1+switch_{overhead})+ \nonumber \\
    nonswitch_{time}\times(1+nonswitch_{overhead}).\nonumber
\vspace{-1mm}
\end{align}
Let us denote by $x$ the proportion of time spent on context switching during program execution (i.e., $x=\frac{switch_{time}}{switch_{time} + nonswitch_{time}} \in (0,1)$). Then the performance slowdown can be estimated as:
\vspace{-1mm}
\begin{align}
    slowdown &= \frac{program_{time'}-program_{time}}{program_{time}} \nonumber \\
    & =  x \times switch_{overhead} \nonumber \\
    & \quad + (1-x) \times nonswitch_{overhead}. \nonumber
    \vspace{-1mm}
\end{align}
Given the statistics of $switch_{overhead}$ by injected loads in context switches under 0.6\% (Figure~\ref{fig:vli_cdf}) and $nonswitch_{overhead}$ by prefetcher-entry erasure under 0.2\% \cite{chen2023afterimage}, the preceding formulated slowdown is up to 0.6\% in practice.

\begin{figure}[t]
    \centering
    \includegraphics[width=\linewidth]{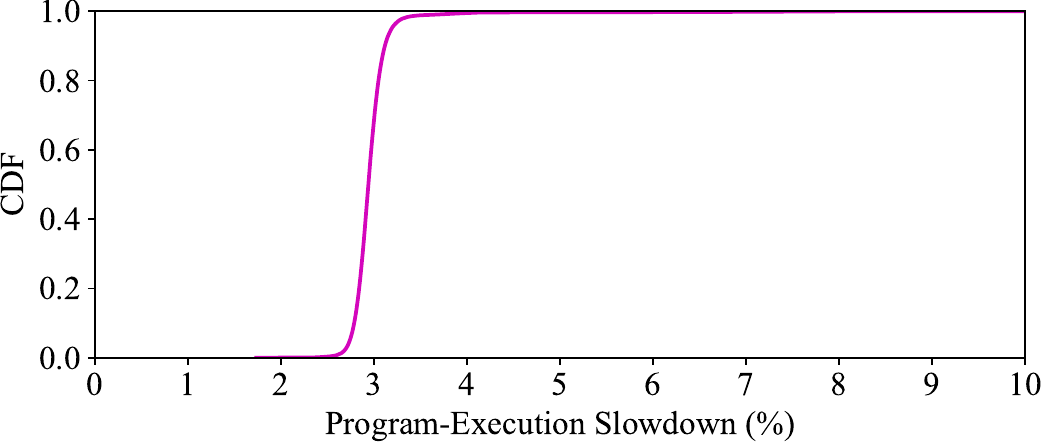}
    \vspace{-6mm}
    \caption{PhantomFetch-vLR overhead in terms of slowdown.}
    \vspace{-6mm}
    \label{fig:vlr_CDF}
\end{figure}

\subsection{PhantomFetch-vLR}

We now present the evaluation results of PhantomFetch-vLR. Since PhantomFetch-vLR operates on secret-dependent blocks, the proportion of such blocks and their invocation frequency may affect performance overhead. In comparison with testing various commonly used benchmarks, we consider it more representative to directly apply PhantomFetch-vLR to real-world RSA \cite{Montgomery-Ladder-RSA}, which is exactly the attacking target in \cite{chen2023afterimage}. Figure~\ref{fig:vlr_CDF} reports the performance overhead in terms of slowdown over 100,000 runs. Almost 99.5\% of the measurements are no more than 4.0\%.





\section{Conclusion}
\label{sec:conclusion}

We have presented and evaluated PhantomFetch as the first prefetching-retentive and hardware-agnostic defense against side-channel attacks exploiting the IP-stride prefetcher. Such attacks essentially exploit the coupling of trained prefetcher entries with secret-dependent loads by the victim. PhantomFetch breaks the exploitable coupling by obfuscating the sensitive load effects of the victim. We have explored two run-time obfuscation schemes---load injection and load relocation. Load injection injects a set of well-crafted loads to erase trained prefetcher entries by the victim before any other process can access and exploit these entries. Load relocation randomly swaps locations of the victim's secret-dependent loads such that the same trained prefetcher entry could couple with any secret-dependent load. 
The experiment results show that they can secure prefetching with only negligible overhead.



\bibliographystyle{plain}
\bibliography{reference}



%

\end{document}